\documentclass[]{aa}

\usepackage{graphicx}
\usepackage{psfig}

\def\til{$\sim$\thinspace}
\def\msun{${\rm M}_{\odot}$}

\def\deg{$^{\circ}$}

\def\ha{H$\alpha$\space}

\def\her{He\thinspace{\sc i}\thinspace$\lambda{6678}$\space}
\def\hher{He\thinspace{\sc i}\thinspace$\lambda{6678}$}
\def\he{He\thinspace{\sc i}\thinspace$\lambda{4471}$\space}

\def\th{\thinspace}

\begin{document}

\title{A new mass-ratio for the X-ray Binary X2127+119 in M15?}

\titlerunning{A new mass-ratio for X2127+119?}

\author{L. van Zyl\inst{1,2}\thanks{E-mail: lvz@astro.keele.ac.uk} \and Z. Ioannou\inst{1,5} \and P.A. Charles\inst{2,3} \and T. Naylor\inst{1,4}}

\authorrunning{L. van Zyl et al.}

\institute{Astrophysics Group, School of Chemistry and Physics,
           Keele University, Staffordshire, ST5 5BG \and
           Department of Astrophysics, Oxford University, Keble Road, 
           Oxford OX1 3RH \and
           Department of Physics \& Astronomy, University of
           Southampton, Southampton SO17 1BJ \and
           School of Physics, University of Exeter, Stocker Road, Exeter EX4 4QL \and
           Department of Astronomy, University of Texas at Austin, Austin, TX
           78712, USA
       }
\date{Received / Accepted }

\abstract{The luminous low-mass X-ray binary X2127+119 in the core of the globular cluster M15 (NGC 7078), which has an orbital period of 17 hours, has long been assumed to contain a donor star evolving off the main sequence, with a mass of 0.8~\msun\space (the main-sequence turn-off mass for M15). We present orbital-phase-resolved spectroscopy of X2127+119 in the \ha and \her spectral region, obtained with the Hubble Space Telescope. We show that these data are incompatible with the assumed masses of X2127+119's component stars. The continuum eclipse is too shallow, indicating that much of the accretion disc remains visible during eclipse, and therefore that the size of the donor star relative to the disc is much smaller in this high-inclination system than the assumed mass-ratio allows. Furthermore, the flux of X2127+119's \her emission, which has a velocity that implies an association with the stream-disc impact region, remains unchanged through eclipse, implying that material from the impact region is always visible. This should not be possible if the previously-assumed mass ratio is correct. In addition, we do not detect any spectral features from the donor star, which is unexpected for a 0.8-\msun\space sub-giant in a system with a 17-hour period.

\keywords{Accretion -- Stars: X-ray binaries -- Stars: individual: X2127+119, AC211}}

\maketitle

\section{Introduction}

\begin{figure*}
\centering
\includegraphics[bb=51 473 562 740,width=\textwidth]{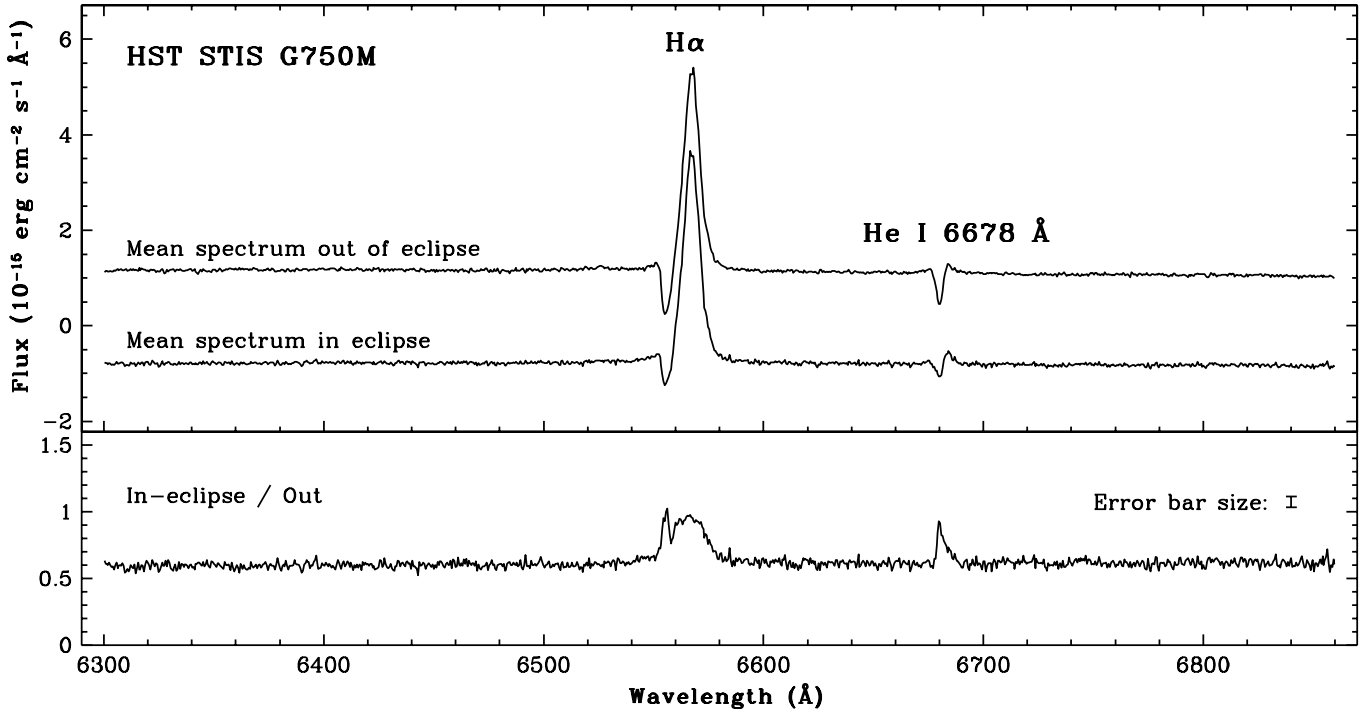}
\vspace{5mm}
\caption{Mean spectra of AC211 in and out of eclipse (orbital-motion-corrected averages of 3 and 12 spectra respectively), obtained with the HST/STIS (see section 2 for details). The mean eclipse spectrum, in which the continuum flux is $0.71 \times 10^{-15} {\rm erg~cm}^{-2}~{\rm s}^{-1}~{\rm \AA}^{-1}$, has been offset for clarity. The error bars for both spectra are smaller than the line thickness. The bottom panel is the eclipse spectrum divided by the out-of-eclipse spectrum, which reveals no evidence of the companion star.}\label{fig:eclipse}
\end{figure*} 

The X-ray binary X2127+119 (AC211) in the core of the globular cluster M15 is optically one of the brightest and most frequently observed of the low-mass X-ray binaries (LMXBs), and yet it remains a highly enigmatic system. Observationally the system presents a considerable challenge -- even under the best seeing conditions spectra of AC211 are heavily contaminated by light from the dense cluster core. AC211's optical and X-ray light curves, along with its very low X-ray-to-optical luminosity ratio, reveal it to be a classic accretion disc corona (ADC) source (Fabian, Guilbert \& Callanan 1987; Naylor et al. 1988). ADC sources are LMXBs viewed at high inclinations, so that the compact object and hot, luminous inner disc are obscured by the accretion disc rim. The X-ray flux we observe is a result of photon-scattering by a large corona above the disc, and is only a small fraction of the source's intrinsic X-ray flux. Indeed the X-ray flux from AC211 has been recently shown to be \til 2-3 times less than previously thought, with the discovery (White \& Angelini 2001) that X2127+119 actually consists of two X-ray binaries within the core of M15. Type I X-ray bursts previously believed to have come from AC211 (Dotani et al. 1990) are now assumed to have been produced by the second X-ray binary.

Ground-based spectroscopy of AC211 reveals evidence for complex and structured mass outflow in the Balmer and \he line profiles (Naylor et al. 1988, 1989; Ilovaisky 1989). Although the contamination problems have made interpretation of the line profiles difficult, as a result of this work it has for a long time been believed that AC211 is in an unusual evolutionary state (Bailyn, Garcia \& Grindlay 1989), in which the system is thought to have a common envelope (although recent work may cast doubt on this [van Zyl et al. 2002]).

Little is known about AC211's compact object, but it has been assumed (e.g. by Homer \& Charles 1998; Ioannou et al. 2002, 2003) to be a neutron star: no globular cluster X-ray binary is known to contain a black hole, and neutron stars are abundant in globular clusters (Hut, Murphy \& Verbunt 1991). The system has an orbital period of 17.1 hours, which makes it unlikely that the companion star is on the main sequence (at an orbital period of 17.1 hours a main-sequence donor star would not fill its Roche lobe). The main-sequence turn-off point for M15 occurs at \til 0.8 M$_{\odot}$ (Fahlman et al. 1985). Therefore it has been assumed that AC211's companion star is an 0.8-M$_{\odot}$ star which has relatively recently begun to evolve off the main sequence (e.g. Bailyn \& Grindlay 1987; Homer \& Charles 1998; Ioannou et al. 2002, 2003).

\section{Observations}

We used the Space Telescope Imaging Spectrograph (STIS) on board the Hubble Space Telescope to obtain orbital-phase-resolved spectra of the \ha and \her region. We used the G750M grating, which gave a spectral range of 6295 -- 6866 \AA\space and a dispersion of 0.56 \AA~pix$^{-1}$. We obtained a total of 15 \til 900-s exposures over 5 HST orbits, each exposure spanning \til 0.015 in the binary's orbital phase.

We timed the observations to correspond to phases \til0.7 to \til1.2 of the binary orbit (calculated according to the latest ephemeris by Ioannou et al. [2003]), in order to follow the behaviour of the lines through eclipse. In addition, with the luminous inner accretion disc obscured during eclipse, we hoped this would provide an opportunity to reveal the spectrum of the companion star. The spectra extracted by STScI's automatic pipeline process had a poor signal-to-noise ratio, so we re-extracted the spectra from the flat-fielded CCD images using T. Marsh's {\sc pamela} spectral reduction software, which applies an optimal extraction technique (Horne 1986).

The spectra are featureless apart from the presence of the \ha and \her lines, both of which show P-Cyg profiles (Fig.~\ref{fig:eclipse}). In the following subsections we describe the behaviour of the continuum and the two spectral features.

\subsection{The elusive companion star}

We had hoped to detect spectral lines from AC211's companion star. Determining its spectral type might enable us to constrain the parameters of the system, and examine its evolutionary state. If the companion's spectrum reveals an atmosphere with a very low gravity, for example, this might indicate that AC211 is a common envelope system, as proposed by Bailyn, Garcia \& Grindlay (1989). However, we found no evidence of spectral lines from the companion star. This was a surprise, because a sub-giant (as the donor is believed to be) in a system with a 17-hour orbital period would be expected to be sufficiently luminous to show some spectral features.

Fig.~\ref{fig:eclipse} compares the mean spectrum in eclipse to the mean spectrum out of eclipse. There is no evidence of a cool stellar spectrum, in particular the 6495-\AA\space feature (a blend of Ca{\th}{\sc i} and Fe{\th}{\sc i} which is the signature of a late G- or K-type star), and the TiO band at 6150-6350~\AA\space (a signature of an M-type star). 

These features are also absent in the mean spectra velocity-corrected\footnote{Correcting for binary's orbital motion, assuming an inclination of 90\deg\space and a compact-object mass of 1.4~\msun.} for companion masses ranging from 0.8 to 0.1~\msun. We tried cross-correlating the continuum regions of interest with combinations of individual spectra and sums of spectra for different assumed companion masses. The resulting cross-correlation functions were noisy, and the main peak was never found to move in a systematic way with phase.

\begin{figure} 
\centering
\includegraphics[bb=52 71 328 224,width=8cm]{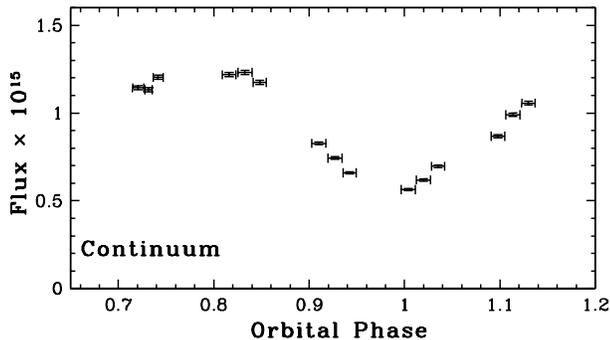}
\caption{Continuum flux through eclipse, measured from our HST/STIS spectra.}\label{fig:conteclipse}
\end{figure} 

\subsection{The continuum eclipse}

In Fig.~\ref{fig:conteclipse} we plot the flux of the continuum between 6300 and 6850 \AA\space against orbital phase. The continuum flux is clearly obscured as the companion star passes in front of the accretion disc, as is expected from AC211's high inclination. The continuum flux out of eclipse is \til$1.2 \times 10^{-15}\;{\rm erg~cm}^{-2}~{\rm s}^{-1}~{\rm \AA}^{-1}$. At mid-eclipse it drops to \til$0.7 \times 10^{-15}\;{\rm erg~cm}^{-2}~{\rm s}^{-1}~{\rm \AA}^{-1}$, or \til60\% of the flux out of eclipse.

\subsection{The \ha and \her lines}

The \ha emission shows no orbital dependence in either flux or velocity: the line flux remains unchanged in eclipse, even though the continuum flux is strongly affected. The \ha absorption component's velocity also shows no variation with phase, but its flux decreases by a factor of \til2 in eclipse.

\begin{figure*} 
\centering
\includegraphics[angle=-90,width=\textwidth]{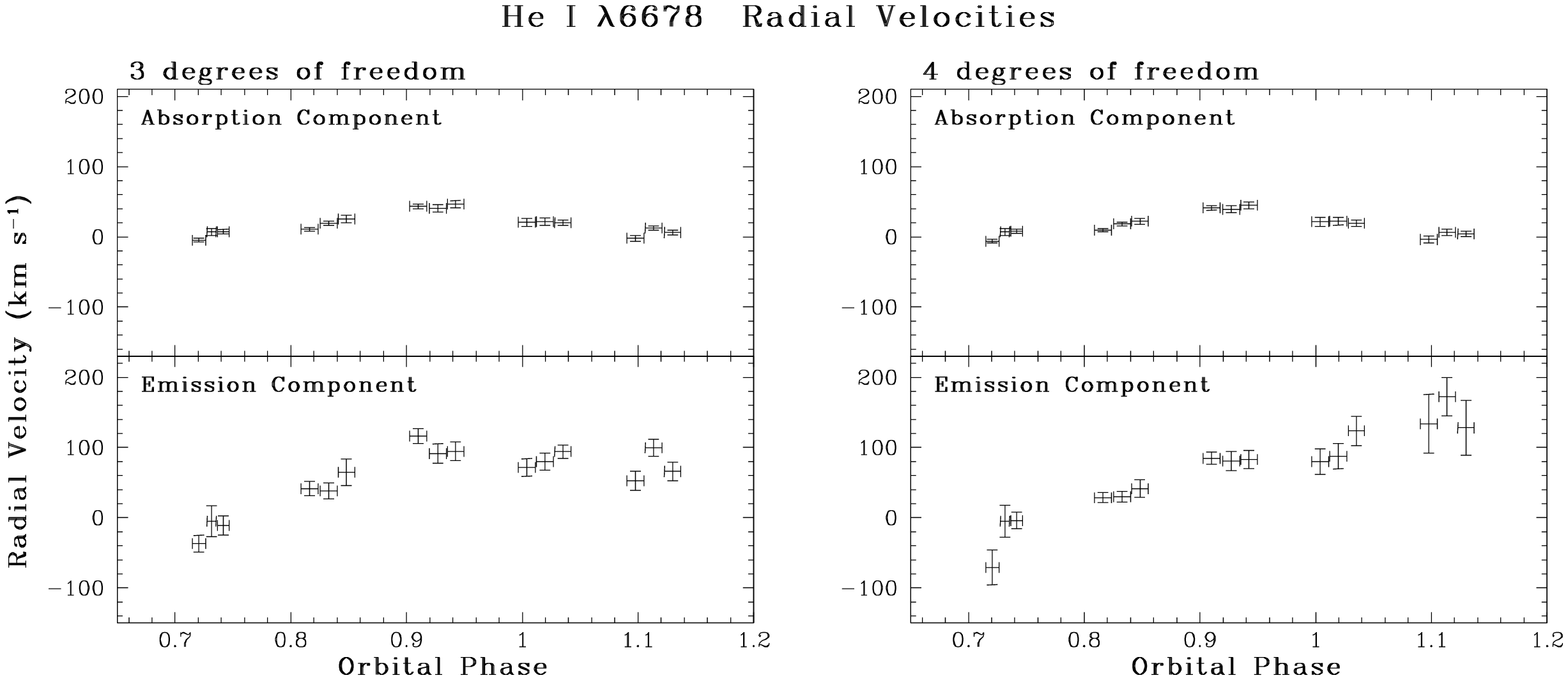}
\vspace{5mm}
\caption{Radial velocities of the \her P Cyg components.}\label{fig:vel}
\end{figure*}

The \her lines consist of a broad emission component with a superimposed narrow absorption component (Figs.~\ref{fig:eclipse} \& \ref{fig:wires}). The velocity of the \her absorption (Fig.~\ref{fig:vel}) shows little variation through eclipse. The behaviour of the \her emission component is much more dramatic. From orbital phases 0.7 to 1.1 the emission moves from blue to red, relative to the cluster's velocity.

Velocities were determined by fitting a pair of Gaussians to each profile, simultaneously fitting the emission and absorption features. In doing this we made the assumption that the emission and absorption components do not interact (and therefore can be added), which may or may not be correct. In Tables 1 and 2 we show the results of double-Gaussian fitting to the emission and absorption components of the \her P-Cyg line profiles, allowing 4 and 3 degrees of freedom. Fitting double-Gaussians with 6 degrees of freedom gave solutions that were not very stable. Therefore we kept the Gaussian widths fixed at the widths of the line components in the mean spectrum, and allowed the velocities and flux heights of the components to vary. In the case of Table 2 (3 degrees of freedom) we kept the height of the emission component fixed to the emission height in the mean specrum. Although it is not necessarily correct to assume that the widths and emission height of the line components stay constant, it is necessary to make these assumptions in order to constrain the fits to give physically realistic solutions. As Tables 1 and 2 show, there are differences in the solutions for different degrees of freedom, but the shape and general trends of the radial velocity curves (Fig.~\ref{fig:vel}) are consistent. (In the tables, the velocities are with respect to the cluster velocity for \hher.)

\begin{table*}
\caption{Results of double-Gaussian fits to the \her P-Cyg profiles: 4 degrees of freedom. In this set of fits, the Gaussian widths are kept fixed to the mean widths of the absorption and emission components, and the continuum flux for each profile is kept fixed at the continuum for each spectrum. The line components' heights (with respect to the continuum levels) and velocities (with respect to the cluster velocity) are allowed to vary.}
\begin{tabular}{cccccc}
\hline
Orbital & V$_{_{\rm Emission}}$ & V$_{_{\rm Absorption}}$ & h$_{_{\rm Emission}}$ & h$_{_{\rm Absorption}}$ & $\chi^2$ \\
Phase   & (km~s$^{-1}$) & (km~s$^{-1}$) & {\small (${\rm erg\,cm}^{-2}{\rm s}^{-1}{\rm \AA}^{-1}$)} & {\small (${\rm erg\,cm}^{-2}{\rm s}^{-1}{\rm \AA}^{-1}$)} & \\
\hline
0.721 & -75.6  $\pm$ 49.8 &   -6.1  $\pm$  5.4 &  0.29   $\pm$ 0.05 &  -1.03   $\pm$ 0.07 &  1.649 \\
0.732 & -10.0  $\pm$ 46.0 &    7.0  $\pm$ 10.2 &  0.43   $\pm$ 0.09 &  -1.25   $\pm$ 0.11 &  0.941 \\
0.742 &  -8.8  $\pm$ 23.6 &    8.2  $\pm$  6.2 &  0.52   $\pm$ 0.05 &  -1.33   $\pm$ 0.06 &  1.287 \\
0.816 &  23.7  $\pm$ 14.2 &    9.4  $\pm$  4.8 &  0.66   $\pm$ 0.04 &  -1.35   $\pm$ 0.05 &  2.909 \\
0.833 &  24.9  $\pm$ 14.8 &   18.4  $\pm$  5.2 &  0.69   $\pm$ 0.04 &  -1.39   $\pm$ 0.05 &  0.654 \\
0.848 &  36.5  $\pm$ 25.2 &   22.2  $\pm$  8.6 &  0.68   $\pm$ 0.08 &  -1.36   $\pm$ 0.09 &  0.796 \\
0.910 &  79.6  $\pm$ 17.4 &   41.1  $\pm$  6.2 &  0.65   $\pm$ 0.05 &  -1.09   $\pm$ 0.06 &  1.381 \\
0.927 &  75.6  $\pm$ 27.6 &   39.4  $\pm$  9.8 &  0.51   $\pm$ 0.06 &  -0.87   $\pm$ 0.07 &  0.906 \\
0.942 &  78.0  $\pm$ 25.4 &   44.9  $\pm$  9.6 &  0.51   $\pm$ 0.06 &  -0.85   $\pm$ 0.07 &  1.024 \\
1.004 &  75.0  $\pm$ 36.0 &   21.5  $\pm$ 13.0 &  0.38   $\pm$ 0.06 &  -0.55   $\pm$ 0.07 &  0.761 \\
1.020 &  82.5  $\pm$ 35.8 &   22.2  $\pm$ 11.0 &  0.39   $\pm$ 0.05 &  -0.61   $\pm$ 0.07 &  0.980 \\
1.035 & 118.7  $\pm$ 41.8 &   19.4  $\pm$  9.0 &  0.32   $\pm$ 0.04 &  -0.51   $\pm$ 0.07 &  1.435 \\
1.098 & 129.0  $\pm$ 83.6 &   -3.7  $\pm$ 10.0 &  0.20   $\pm$ 0.05 &  -0.63   $\pm$ 0.07 &  1.131 \\
1.114 & 167.4  $\pm$ 54.2 &    6.6  $\pm$  9.4 &  0.25   $\pm$ 0.03 &  -0.66   $\pm$ 0.05 &  1.149 \\
1.130 & 123.0  $\pm$ 77.4 &    4.4  $\pm$  7.8 &  0.22   $\pm$ 0.05 &  -0.76   $\pm$ 0.08 &  1.126 \\

\hline
\end{tabular}
\end{table*}

\begin{table*}
\caption{Results of double-Gaussian fits to the \her P-Cyg profiles: 3 degrees of freedom. In this set of fits, the Gaussian widths are kept fixed to the mean widths of the absorption and emission components, and the continuum flux for each profile is kept fixed at the continuum for each spectrum. In addition, the height of the emission component is kept fixed at the height of the emission component in the summed spectrum. The absorption components' height and the line components' velocities (with respect to the cluster velocity) are allowed to vary.}
\begin{tabular}{ccccc}
\hline
Orbital & V$_{_{\rm Emission}}$ & V$_{_{\rm Absorption}}$ & h$_{_{\rm Absorption}}$ & $\chi^2$ \\
Phase   & (km~s$^{-1}$) & (km~s$^{-1}$) & {\small (${\rm erg\,cm}^{-2}{\rm s}^{-1}{\rm \AA}^{-1}$)} & \\
\hline
0.721 & -41.9  $\pm$ 23.8 &   -4.5  $\pm$  5.4 &   -1.22  $\pm$ 0.01    & 1.649 \\
0.732 &  -9.9  $\pm$ 44.4 &    7.0  $\pm$ 10.2 &   -1.25  $\pm$ 0.02    & 0.941 \\
0.742 & -16.0  $\pm$ 27.2 &    7.5  $\pm$  6.4 &   -1.22  $\pm$ 0.01    & 1.287 \\
0.816 &  36.3  $\pm$ 20.4 &   10.7  $\pm$  5.6 &   -1.09  $\pm$ 0.01    & 2.909 \\
0.833 &  33.2  $\pm$ 23.2 &   19.5  $\pm$  6.4 &   -1.09  $\pm$ 0.01    & 0.654 \\
0.848 &  59.7  $\pm$ 37.4 &   25.4  $\pm$ 10.2 &   -1.06  $\pm$ 0.02    & 0.796 \\
0.910 & 111.2  $\pm$ 21.0 &   43.4  $\pm$  6.8 &   -0.80  $\pm$ 0.02    & 1.381 \\
0.927 &  86.5  $\pm$ 27.6 &   40.6  $\pm$ 10.2 &   -0.77  $\pm$ 0.02    & 0.906 \\
0.942 &  89.6  $\pm$ 26.2 &   46.6  $\pm$ 10.0 &   -0.75  $\pm$ 0.02    & 1.024 \\
1.004 &  66.6  $\pm$ 25.0 &   20.6  $\pm$ 11.8 &   -0.62  $\pm$ 0.02    & 0.761 \\
1.020 &  74.8  $\pm$ 23.8 &   21.5  $\pm$ 10.2 &   -0.66  $\pm$ 0.02    & 0.980 \\
1.035 &  89.0  $\pm$ 19.4 &   20.0  $\pm$  7.8 &   -0.66  $\pm$ 0.02    & 1.435 \\
1.098 &  47.7  $\pm$ 27.2 &   -2.0  $\pm$  8.0 &   -0.95  $\pm$ 0.02    & 1.131 \\
1.114 &  94.8  $\pm$ 23.8 &   12.5  $\pm$  6.2 &   -0.92  $\pm$ 0.02    & 1.149 \\
1.130 &  60.9  $\pm$ 26.2 &    6.1  $\pm$  6.8 &   -1.04  $\pm$ 0.02    & 1.126 \\
\hline
\end{tabular}
\end{table*}

\section{Discussion}

\subsection{The \ha line profiles}

The strong \ha P Cyg profiles suggest an outflow arising from the ADC or accretion disc wind. Compared to winds in other interacting binaries (e.g. C\'{o}rdova 1995), this wind is very slow. Measuring the extent of the blue-shifted component of \ha gives a line-of-sight velocity of only $600 \pm 25$ km~s$^{-1}$, perhaps indicating that we are viewing a strongly bipolar disc wind edge-on.

\subsection{The \her line profiles}

\begin{figure} 
\centering
\includegraphics[width=8.3cm]{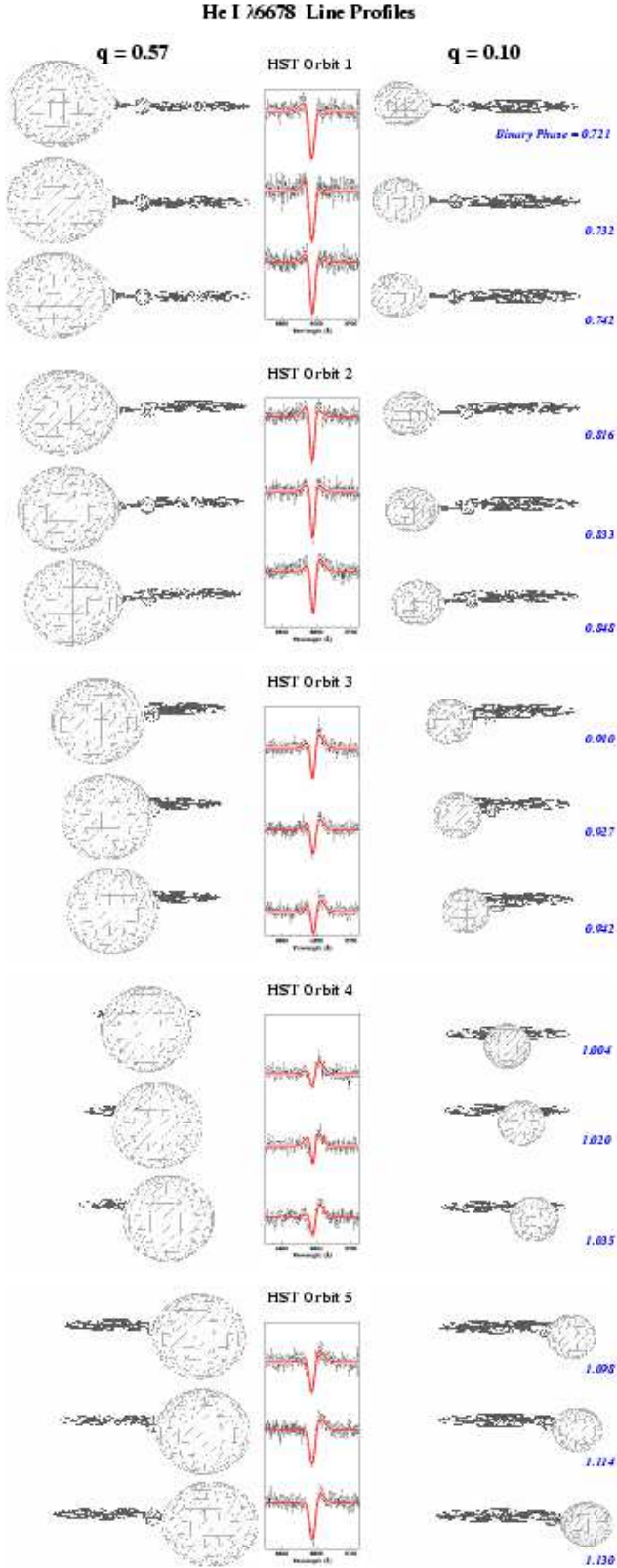}
\vspace{4mm}
\caption{The \her line profiles (the solid lines superimposed on each spectrum are the best-fitting profiles modelled by pairs of Gaussians) with simple representations of an LMXB plotted alongside for each orbital phase. Two sets of cartoons are calculated, using mass ratios M$_2$/M$_{\rm x}$ of 0.57 (left) and 0.10 (right). The binary inclination assumed is 84\deg. The cartoons are simple and do not adequately represent the stream/disc impact region and disc bulge, which in reality expends along the disc rim for almost 90\deg\space ahead of the stream/disc impact site.}\label{fig:wires}
\end{figure}

In Fig.~\ref{fig:wires} we show the relationship between the \her line profiles and the geometry of the binary system for two different mass ratios (defined as $q=M_{2}/M_{\rm x}$). The cartoons are computed assuming 1.4~\msun\space for the compact object, 0.8~\msun\space (left) or 0.1~\msun\space (right) for the companion, an inclination of 84\deg\space (from X-ray and UV light curves, AC211's inclination is $>$80\deg\space [e.g. Ilovaisky et al. 1993; Ioannou et al. 2002]). The location of the stream-disc impact site is indicated. What is not shown is the luminous, thick bulge on the accretion disc rim which begins at the stream-disc impact site and which extends along the disc rim almost 90\deg\space ahead of the impact site (Ioannou et al. 2002).

From this figure and the radial velocity curve (Fig.\ref{fig:vel}), the \her emission component appears to be associated with the motion of the accretion stream or the stream-disc impact site: the radial velocity curve of the \her emission is clearly asymmetric, and initially blue-shifted at orbital phase \til0.7, it rises to redder velocities until it reaches a plateau at orbital phase \til0.9, the phase near which the accretion stream and stream-disc impact region are moving parallel to our line-of-sight and receding from us.

The flux of the \her emission remains constant through eclipse, indicating that the \her emission-line-forming region is very extended. For disc-bulge splash to be visible throughout eclipse implies either that the system's inclination is lower than has been assumed, or that the companion star is very much smaller, in relation to the disc bulge, than has been assumed.

We know that AC211 has a high inclination: its X-ray and optical light-curves show deep dips caused by disc-rim obscuration of the corona and inner disc (Ilovaisky et al. 1993; Ioannou et al. 2002), and the compact object is always obscured. Since LMXB disc opening (semi-)angles are \til10\deg\space (e.g. de Jong et al. 1996), AC211 must therefore have an inclination of at least 80\deg\space to keep the compact object obscured at all orbital phases, and from its large amplitude dipping, it possibly has an inclination significantly greater than 80\deg\space (Ioannou et al. 2002). Therefore, the visibility of \her throughout eclipse suggests that the companion star is small relative to the disc and disc bulge, requiring that the system has a significantly smaller mass ratio than previously believed.

It has been argued (for example by Drew \& Verbunt [1985] in the case of UX UMa and RW Tri) that the absence of eclipses of P-Cyg lines can be due to a simultaneous eclipse of both the emission and absorption components, and therefore does not necessarily imply an extreme mass ratio. However, we believe that this does not apply in the case of AC211's \her P-Cyg lines, as at least part of the flux is contributed by an accretion stream, not a disc wind.

\subsection{The Continuum Eclipse}

Further evidence for a significantly smaller mass ratio comes from the continuum eclipse (Fig.~\ref{fig:conteclipse}). The continuum flux in eclipse drops to \til60\% of the flux out of eclipse, indicating that there is a substantial uneclipsed flux from the accretion disc which dominates the light from the system between 6300--6850~\AA\space during eclipse. How much of this flux could be contributed by the companion star? If we assume the companion is an 0.8~\msun\space star just beginning to evolve off the main sequence, then it has a V-magnitude of ~\til18.5 (from Durrell \& Haris [1993]), and contributes \til5~\% of the total light from the system in the V-band (AC211 has a V-magnitude of \til15.5). If we assume that the companion's R-band contribution is not too dissimilar, then it contributes a very small fraction to the continuum flux between 6300--6800~\AA\space during the eclipse, nowhere near the \til60\% of the out-of-eclipse flux level that we observe. Therefore, a substantial part of the disc remains unobscured during eclipse, again requiring that the system has a significantly smaller mass ratio than previously believed.

\subsection{The Binary Parameters}

Is it possible to use these data to constrain the binary parameters? We discuss some possibilities below.

In the cartoons plotted on the left-hand side of Fig.~\ref{fig:wires} we assume the previously assumed values for the masses of AC211's components, namely \til 1.4~\msun\space for the compact object (based on the fact that all other globular cluster LMXBs contain neutron stars) and \til 0.8~\msun\space for the companion star (the main-sequence turn-off mass for M15), and the high inclination generally assumed for AC211 (inferred from the observed optical and X-ray light curves). It is immediately clear that this picture is incompatible with our STIS spectra: in the cartoons, the disc is almost completely obscured during eclipse, while in our spectra the continuum flux (assumed to come mainly from the accretion disc) drops by only \til 40\% (Fig.~\ref{fig:conteclipse}).

This leads us to consider the possibility of a substantially lower mass ratio. As shown on the right-hand side of Fig.~\ref{fig:wires}, this can neatly solve the problem of the shallow continuum eclipse without lowering the inclination: most of the (outer) disc remains visible during eclipse. If AC211's mass ratio is indeed much lower than the generally accepted value of 0.57, then either the companion has a very low mass, \til0.1~\msun, or the compact object is very massive: \til8~\msun.

If the companion has a mass of \til0.1~\msun, it would need to be an evolved star. It would initially have been a more massive star ($\geq$ 0.8~\msun) which has shed almost all of its H-rich envelope. A test for this scenario would be to look for unusual chemical abundances and nucleosynthesis products (He-enrichment, CNO-processed material) in AC211's spectrum. This scenario has been proposed before for LMXBs with low-mass donors (Webbink, Rappaport \& Savonije 1983).

The major drawback of a very-low-mass, evolved star scenario for AC211 is that the standard model predicts that the mass-transfer rate would be substantially lower than the \til 10$^{-7}-10^{-8}$ M$_{\odot}$~y$^{-1}$ inferred from AC211's X-ray luminosity (Podsiadlowski, Rappaport \& Pfahl 2001). It is possible for stripped, evolved companions to produce these high mass-transfer rates, but only if the onset of mass transfer occurs after the companion has become a giant, in which case the initial binary orbital period exceeds \til 50 days, and increases as the system evolves (McMillan, Taam \& McDermott 1990). In order to produce orbital periods of \til 17~h, the onset of mass transfer must occur while the companion is still a sub-giant (Pylyser \& Savonije 1988).

A possible way to increase the mass-transfer rate in this scenario is through angular momentum loss from the system, which can substantially increase mass-transfer rates in extreme-mass-ratio systems (Taam \& Wade 1985). The evidence for mass outflows from AC211 in studies of \he (Naylor et al. 1988; Ilovaisky 1989; van Zyl et al. 2002) suggests that this mechanism cannot be ruled out for AC211. 

Another way to increase the mass-transfer rate is through irradiation of the companion (Podsiadlowski 1991). With such a small mass ratio, the companion will be well shielded from the compact object by the disc rim, but it would still be exposed to X-ray irradiation from the ADC. If the ADC scatters \til 10$^{-4}$ to \til 10$^{-3}$ of the system's X-ray flux onto the companion, that would be enough to completely change its structure and drive up the mass-transfer rate (Podsiadlowski 2001, private communication).

\section{Conclusions}

We have presented the first orbital-phase-resolved spectroscopy of X2127+119 in the \ha and \her spectral region, and we show that these data are incompatible with AC211's assumed mass-ratio of 0.57. The continuum eclipse is too shallow, indicating that much of the accretion disc remains visible during eclipse, and therefore that the size of the donor star relative to the disc is much smaller in this high-inclination system than the assumed mass-ratio allows. Furthermore, the flux of X2127+119's \her emission, which has a velocity that implies an association with the stream-disc impact region, remains unchanged through eclipse, implying that material from the impact region is always visible. This should not be possible if the previously-assumed mass ratio is correct. In addition, we do not detect any spectral features from the donor star, which is unexpected for a 0.8~\msun\space sub-giant in a system with a 17-hour period.

Our results indicate that AC211 has a significantly smaller mass ratio, closer to \til0.1. This implies that either the compact object is very massive (up to 8~\msun), or that the companion star has a very low mass (\til0.1~\msun).

\section*{Acknowledgments}

Based on observations made with the NASA/ESA Hubble Space Telescope, obtained at the Space Telescope Science Institute, which is operated by the Association of Universities for Research in Astronomy, Inc., under NASA contract NAS 5-26555. These observations are associated with proposal number 8343.

LvZ acknowledges the hospitality of the University of Southampton, where much of this paper was written. LvZ would like to thank Rob Hynes for the use of his {\sc binsim} binary visualization code, Tom Marsh for the use of his {\sc molly} and {\sc pamela} spectral analysis software, and Philipp Podsiadlowski and Christian Knigge for helpful discussions. LvZ acknowledges the support of scholarships from the Vatican Observatory, the National Research Foundation (South Africa), the University of Cape Town, and the Overseas Research Studentship scheme (UK).

\label{lastpage}

\end{document}